\documentclass[fleqn,usenatbib]{mnras}

\usepackage{newtxtext}
\usepackage[T1]{fontenc}

\DeclareRobustCommand{\VAN}[3]{#2}
\let\VANthebibliography\thebibliography
\def\thebibliography{\DeclareRobustCommand{\VAN}[3]{##3}\VANthebibliography}

\usepackage{graphicx}
\usepackage{natbib}
\usepackage{amssymb}
\usepackage{amsmath}
\usepackage{xspace}
\usepackage[dvipsnames]{xcolor}
\usepackage{dirtytalk}
\usepackage{placeins}

\usepackage[normalem]{ulem}
\usepackage{soul}
\usepackage[varg]{txfonts}

\newcommand{\gaga}{$\gamma\gamma$}

\newcommand{\rb}[1]{{\color{black}  #1}} 
\newcommand{\is}[1]{{\color{black}  #1}} 

\graphicspath{{./}{figures/}}

\title[Very young SNRs in dense environments]{Core-collapse supernovae in dense environments - particle acceleration and non-thermal emission}

\author[Brose et al.]{R. Brose$^{1}$\thanks{E-Mail: broserob@cp.dias.ie} 
    I. Sushch$^{2,3}$
    J. Mackey$^{1}$
\\
$^1$Dublin Institute for Advanced Studies, Astronomy \& Astrophysics Section,  DIAS Dunsink Observatory, Dublin D15 XR2R, Ireland\\ 
$^2$ Centre for Space Research, North-West University, 2520 Potchefstroom, South Africa\\
$^3$Astronomical Observatory of Ivan Franko National University of Lviv, Kyryla i Methodia 8, 79005 Lviv, Ukraine}

\date{Received ; accepted}

\pubyear{2022}

\begin{document}
\label{firstpage}
\pagerange{\pageref{firstpage}--\pageref{lastpage}}
\maketitle



\begin{abstract}
Supernova remnants are known to accelerate cosmic-rays from the detection of non-thermal emission in radio waves, X-rays, and gamma-rays. However, the ability to accelerate cosmic-rays up to PeV energies has yet to be demonstrated. The presence of cut-offs in the gamma-ray spectra of several young SNRs led to the idea that PeV energies might only be achieved during the first years of a remnant’s evolution. We use our time-dependent acceleration-code RATPaC to study the acceleration of cosmic-rays in supernovae expanding into dense environments around massive stars. We performed spherically symmetric 1-D simulations in which we simultaneously solve the transport equations for cosmic-rays, magnetic turbulence, and the hydrodynamical flow of the thermal plasma in the test-particle limit. We investigated typical CSM parameters expected around RSG and LBV stars for freely expanding winds and accounted for the strong $\gamma\gamma$ absorption in the first days after explosion. The maximum achievable particle energy is limited to below  $600\,$TeV even for largest considered values of the magnetic field and mass-loss rates.
The maximum energy is not expected to surpass $\approx200\,$TeV and $\approx70\,$TeV for LBVs and RSGs that experience moderate mass-loss prior to the explosion. We find gamma-ray peak-luminosities consistent with current upper limits and evaluated that current-generation instruments are able to detect the gamma-rays from Type-IIP explosions at distances up to $\approx60\,$kpc and Type-IIn explosions up to $\approx1.0\,$Mpc. We also find a good agreement between the thermal X-ray and radio synchrotron emission predicted by our models with a range of observations.


\end{abstract}
 



\section{Introduction}
Supernova remnants (SNRs) are considered among the best candidates for the source of the Galactic Cosmic Rays (CRs) \citep{1934PNAS...20..259B, 1987PhR...154....1B}. However, evidence that they are able to accelerate CRs up to the required $3\,$PeV is elusive. Observational constraints put the maximum achievable energy at $\approx100\,$TeV at most, whereas young SNRs such as Tycho and Casiopeia A show even lower cutoff energies despite their youth \citep{2020ApJ...894...51A}.

Recent studies of CR acceleration suggest that the escape of CRs once SNRs enter the Sedov stage might be responsible for the observed, soft late-time spectra of evolved SNRs \citep{2019MNRAS.490.4317C, 2020A&A...634A..59B}. The historical remnants Casiopeia A and Tycho are close to entering the Sedov-phase, which makes an acceleration of CRs beyond PeV energies unlikely in the future, as the rapidly decreasing maximum energy is the main driver of the particle escape after the onset of the Sedov phase.

\citet{MurThoLac11} argued that the collision between a supernova blastwave and dense circumstellar material can be an effective site of particle acceleration up to very high energies (1\,PeV and beyond), and predicted that high-energy radiation from such supernovae could be detectable with future observatories from GeV to TeV energies out to distances of up to 200 Mpc in optimistic scenarios \citep{MurThoOfe14}.

\citet{2013MNRAS.431..415B} suggest that the CR density gradient is only large enough during the initial 20 years of a SNRs evolution to drive turbulence to relevant scales (for producing PeV CRs) by the non-resonant streaming instability.
In addition to the gradients resulting from the small shock extension, high-density environments are favorable for the acceleration to PeV energies \citep{2018MNRAS.479.4470M, 2020MNRAS.494.2760C}.
Because of these considerations, it is possible that only a subset of SN explosions harbor the conditions needed to contribute significantly to the Galactic CR flux at PeV energies \citep{2021ApJ...922....7I, 2020APh...12302492C}. 

An observational benefit from SNe exploding in a dense circumstellar-medium (CSM) is that the high density boosts the $\gamma$-ray emission from p-p interactions. Due to the decreasing density of the CSM with increasing distance from the center of explosion, the highest luminosities are achieved shortly after the explosion.
In this case, interactions between the emitted $\gamma$-ray photons and the photons emitted by the SN photosphere attenuate the $\gamma$-ray flux, especially in the first days after the explosion \citep{2020MNRAS.494.2760C, 2022MNRAS.511.3321C}. 

Recently, evidence has been presented of transient $\gamma$-ray emission at \textit{Fermi-LAT} energies overlapping with the positions of SN 2004dj, a Type-IIP explosion \citep{2020ApJ...896L..33X}, and SN iPTF14hls \citep{2018ApJ...854L..18Y}, at distances of $3.5\,$Mpc and $156\,$Mpc, respectively.
Additionally, two more SN candidates have been found in a transient search throughout the \textit{Fermi-LAT} archive \citep{2021MNRAS.tmp.1282P}. So far, there is no sign of a detection of very-high energy (VHE) gamma-ray emission \citep{2019A&A...626A..57H, Abdalla:2021Mg}. 

Since remnants expanding in dense circumstellar media are regarded as the best candidates both for acceleration to PeV energies and for detection of the $\gamma$-ray emission produced by $p$-$p$ collisions, we consider Type-IIn and Type-IIP SNe associated with luminous blue variable (LBV) and red supergiant (RSG) progenitor stars in this work. The mass-loss rate of RSGs (specifically the wind density) is very important for our predictions, but is quite uncertain and the subject of ongoing investigation and revision \citep[e.g.][]{BeaDav18}.
  In this work we consider RSGs with both moderate and high mass-loss rates among the measurements compiled by \citet{MauJos11}.
  We use the term LBV as a SN IIn progenitor very loosely in this work, referring only to a star that has strongly enhanced mass loss shortly before core collapse, producing a CSM much denser and more massive than the CSM around a typical RSG.
  The progenitors of some SN IIn have eruptions and strongly evolving mass-loss rates shortly before core collapse \citep{FraMagKot13, OfeSulSha14}, but it is not clear if the progenitors are actually LBVs \citep{BoiGro18}.
  For example, \citet{SmiHinRyd09} showed that VY CMa, a massive Galactic RSG, would produce a Type IIn SN if it were to explode, and the dense circumstellar shell detected around Betelgeuse \citep{LeBMatGer12} could also produce IIn signatures \citep{MacMohGva14}.
  In this work we consider only smooth constant winds, but we will extend our analysis to time varying winds including eruptions in future work.

This paper is organised as follows:
  section~\ref{sec:methods} introduces our assumptions for density and magnetic field of pre-supernova circumstellar matter, the methods for modelling hydrodynamics, particle acceleration, transport, and energy losses, and our treatment of radiation processes (including the implementation of $\gamma\gamma$ absorption).
  In section~\ref{sec:results} we describe first the thermal X-ray emission predicted from our models in the context of observed supernovae; then the time-evolution of the accelerated particle spectra and predicted $\gamma$-ray emission at GeV and TeV energies.
  The horizon out to which supernovae can be detected by current and future observatories is calculated based on our predictions in section~\ref{sec:gammarays}.
  We also study the predicted non-thermal radio emission in comparison with observations (section~\ref{sec:radio}).
  Section~\ref{sec:conclusions} presents our conclusions.

\section{Basic equations and assumptions}
\label{sec:methods}
In this section, we introduce the numerical methods that were used in this study. The methods presented here a similar to earlier publications using the \textbf{R}adiation \textbf{A}cceleration \textbf{T}ransport \textbf{Pa}rallel \textbf{C}ode (RATPaC). RATPaC was first used to study CR-acceleration (see \ref{sec:CRs}) in Type-Ia SNRs \citep{Telezhinsky.2012a} and \citep{Telezhinsky.2013} assuming Bohm-like diffusion and considering the passive transport of the large-scale magnetic field (see \ref{sec:MF}). Later modules for solving the transport equation of magnetic turbulence \citep{2016A&A...593A..20B} (see \ref{sec:Turb}) and for treating the hydrodynamical evolution on-the-fly were introduced \citep{2018A&A...618A.155S, 2019A&A...627A.166B} (see \ref{sec:Hydro}).

Here we present the basic assumptions we made in our numeric approach to the DSA problem. We combined a kinetic treatment of the CRs with a thermal leakage injection model, a fully time-dependent treatment of the magnetic turbulence, and a \textsc{PLUTO}-based hydro calculation.

\subsection{Circumstellar magnetic field}\label{sec:MF}
We assume that the magnetic field is dynamically unimportant for the hydrodynamial evolution of the SNR. In and around the SNR two components contribute to the magnetic field: a large-scale field shaped for instance by the progenitor's wind and a turbulent component that gets self-amplified by the streaming of CRs. We calculate the total magnetic field strength as 
\begin{align}
    B_\text{tot} &= \sqrt{B_0^2+B_\text{Turb}^2} \text{ , }
\end{align}
where $B_0$ is the large-scale magnetic field and $B_\text{Turb}$ the turbulent component (see section \ref{sec:Turb} for details).

For the large-scale field, we solve the induction equation for a frozen-in field that is transported passively with the hydrodynamic flow \citep{Telezhinsky.2013}. 
Both RSG and LBV stars are expected to have surface magnetic fields which are advected outwards with the stellar wind, even before additional amplification by a turbulent dynamo is considered. The latter might convert a sizable fraction of the bulk flow energy to magnetic field \cite[and references therein]{2018MNRAS.479.4470M}.

There is a rather good understanding of the magnetic field configuration within the free-wind region.
Once the wind becomes super-Alfv\'enic the field lines all become open and
the magnetic field profile can be expressed as \citep{1982ApJ...253..188V}
\begin{equation}
    B_\mathrm{wind}(r) = B_\ast \left[ 1 + \left(\frac{V_\mathrm{w}}{V_\mathrm{rot}}\right)^2 \right]^{-1/2} \frac{R_\ast}{r} \left[ 1 + \left( \frac{R_\ast V_\mathrm{w}}{V_\mathrm{rot}}\right)^{2}\frac{1}{r^2}\right]^{1/2},
\end{equation}
with $B_\ast$ denoting magnetic field strength at the surface of the star, $R_\ast$ the radius of the star, $V_\mathrm{rot}$ rotation velocity of the star, and $V_\mathrm{w}$ velocity of the wind.
The magnetic field is assumed to be frozen into the plasma and thus close to the star it is mostly radial and decreases as $\propto1/r^2$ with the distance from the star, but at larger distances, $r\gtrsim R_\ast V_\mathrm{w}/V_\mathrm{rot}$, it becomes mostly azimuthal due to the stellar rotation and decreases as $\propto1/r$.

For RSGs, with radius $R_\ast\approx5\cdot10^{13}$\,cm, $V_\mathrm{w}\sim (20-50)$\,km\,s$^{-1}$, and $V_\mathrm{rot}\lesssim1$\,km\,s$^{-1}$, this transition happens at $r\gtrsim(1-2.5)\cdot10^{15}$\,cm.
One of the best studied RSGs, Betelgeuse, has a rapid surface-rotation rate of $V_\mathrm{rot}\approx17$\,km\,s$^{-1}$, but this is very unusual and is difficult to explain in the context of single star evolution \citep{2017MNRAS.465.2654W}.
Values of $V_\mathrm{rot}\lesssim1$\,km\,s$^{-1}$ are more in line with theoretical expectations.
LBVs are enigmatic objects, often with bipolar nebulae \citep{2020Galax...8...20W} suggesting rapid rotation during some part of their variability cycle.
Their radii, mass-loss rates and wind speeds all change by a factor of a few on a $\sim10$-year timescale \citep{2021A&A...647A..99G}, but typical wind speed is $V_\mathrm{w}\sim 10^2$\,km\,s$^{-1}$, radius $R_\ast\sim10^{13}$\,cm, and we assume a typical rotation speed $V_\mathrm{rot}\gtrsim0.1V_\mathrm{w}$.
For these values, the transition from radial to toroidal magnetic field occurs at $r\sim10^{14}$\,cm.
If a SN shock is expanding through this medium at $10^4$\,km\,s$^{-1}$, it will cross this transition region after $\sim 1$\,day for a LBV progenitor or $>(10-25)$\,days for a RSG progenitor.


Given these estimates it is unlikely that any observational signature can be detected from the initial expansion in the region where the magnetic field decreases as $r^{-2}$, since radio emission is not able to penetrate the dense CSM early enough, and current-generation $\gamma$-ray telescopes struggle with detections and are unlikely to measure a robust lightcurve.
Hence, we simplified our model by neglecting the initial zone and assumed a magnetic field in the wind, that follows
\begin{align}
    B(r) = B_\ast\frac{R_\ast}{r} \text{ . }
\end{align}
We absorb all the uncertainties of the surface magnetic-field into the variables $B_\ast$ and $R_\ast$. We assumed for both progenitor stars a fixed product of
$B_\ast (R_\ast/R_\odot)=1000$\,G.

There are not many measurements of the surface magnetic field of RSGs: Betelgeuse was measured to have an $\approx 1$\,G field \citep{2010A&A...516L...2A}, and two other RSGs have a similar field strength \citep{2017A&A...603A.129T}.  Given the radii of RSGs ($R_\ast\sim10^3$\,R$_\odot$), these measurements support our choice of $B_\ast$.
For LBVs we know of no detections of surface magnetic fields but, for example, \citet{WhiBreKon20} use $R_\ast=100$\,R$_\odot$ and $B_\ast=100$\,G, for their model of $\eta$ Car, 10 times stronger than the field we assume here.
Additionally, we tested models where we considered a uniform $5\,\mu$G throughout the ambient medium for comparison. 


\subsection{Cosmic rays}\label{sec:CRs}
We use a kinetic approach to model the acceleration of CRs in the test-particle approximation. We made sure that our chosen parameters guarantee a CR-pressure that stays below 10\% of the shock ram-pressure. The time-dependent transport equation for the differential number density of cosmic rays $N$ \citep{Skilling.1975a} is given by:
\begin{align}
\frac{\partial N}{\partial t} =& \nabla(D_r\nabla N-\mathbf{u} N)\nonumber\\
 &-\frac{\partial}{\partial p}\left( (N\dot{p})-\frac{\nabla \cdot \mathbf{ u}}{3}Np\right)+Q
\label{CRTE}\text{ , }
\end{align}
where $D_r$ denotes the spatial diffusion coefficient, $\textbf{u}$ the advective velocity, $\dot{p}$ energy losses and $Q$ the source of thermal particles.

Equation (\ref{CRTE}) is solved using the \textit{FiPy}-library \citep{2009CSE....11c...6G} in a frame co-moving with the shock. The radial coordinate is transformed according to $(x-1)=(x^*-1)^3$, where $x=r/R_{sh}$. This transformation guarantees a very fine resolution close to the shock for a equidistant binning of $x^*$. The outer grid boundary extends to several tens of shock radii upstream for $x^*\gg1$. Thus, all accelerated particles can be kept in the simulation domain. 

\subsubsection{Injection} 
We inject a fixed fraction of particles from the thermal pool as CRs according to the thermal leakage injection model \citep{Blasi.2005a,1998PhRvE..58.4911M}, where the efficiency of injection $\eta_i$ is given by
\begin{align}
\eta_i = \frac{4}{3\sqrt{\pi}}(\sigma-1)\psi^3e^{-\psi^2}\text{ . }
\end{align}
The shock compression-ratio is denoted by $\sigma$, and the multiple of the thermal momentum, at which we inject particles by $\psi$. We use a value of $\psi=4.2$, which seems to be appropriate based on the emission from SN 1006 \citep[see Appendix A]{2021A&A...654A.139B}. We consider a uniform injection efficiency across the entire shock surface. Any effects that can be attributed to the shock obliquity are neglected here \citep[see][section 2.1.1 and references therein]{2021A&A...654A.139B}. 


\subsubsection{Energy losses}
Electrons undergo synchrotron losses in the magnetic field of the SN progenitor's wind, and in the CR-amplified magnetic field close-to and in the downstream of the SN shock. It has to be expected that the inverse-Compton losses are also sizable as long as the emission from the SN-explosion itself is strong. However, \cite{2021ApJ...908...75B} showed that the radio emission for SNe in dense circumstellar environments tends to peak only after the SN emission powered by radioactive decay has faded after hundreds of days \citep[and references therein]{2013A&A...555A..10T, 2013MNRAS.434.1636T}. This initial supression of the radio emission is caused by free-free absorption of radio photons in the dense CSM \citep{2009A&A...499..191T}. Since the inverse-Compton losses became unimportant compared to synchrotron losses by the time the radio emission becomes detectable, we ignored the inverse-Compton losses in this work and are only considering synchrotron losses.  

It has to be noted, that the large magnetic fields result in a very-low energy that is needed by electrons in order to emit at radio frequencies.
The characteristic synchrotron frequency is given by
\begin{align}
    f_c &= 16\,\text{MHz}\left(\frac{B_0}{1\,\mu\text{G}} \right)\left(\frac{E_\text{el}}{5\,\text{GeV}} \right)\label{eq:RadioFc}
\end{align}
Hence, electrons with $2.5\,$MeV emit 8\,GHz radio emission when the field strength is $\approx1\,$G.


\subsection{Magnetic turbulence}\label{sec:Turb}
In parallel to the transport equation for CRs, we solve a transport equation for the magnetic turbulence spectrum, assuming Alfv\'en waves only, and thus calculate the diffusion coefficient self consistently \citep{2016A&A...593A..20B}. In this case, the diffusion coefficient varies strongly in space and time and is coupled to the spectral energy-density per unit logarithmic bandwidth, $E_w$. The evolution of $E_w$ is described by
\begin{align}
 \frac{\partial E_w}{\partial t} +   \nabla \is{\cdot} (\mathbf{u} E_w) + k\frac{\partial}{\partial k}\left( {k^2} D_k \frac{\partial}{\partial k} \frac{E_w}{k^3}\right) = \nonumber\\
=2(\Gamma_g-\Gamma_d)E_w \text{ . }
\label{eq:Turb_1}
\end{align}
Here, $\mathbf{u}$ denotes the advection velocity, $k$ the wavenumber, $D_k$ the diffusion coefficient in wavenumber space, and $\Gamma_g$ and $\Gamma_d$ the growth and damping terms, respectively \citep{2016A&A...593A..20B}.

The diffusion coefficient of CRs is coupled to $E_w$ by 
\begin{align}
    D_{r} &= \frac{4 v}{3 \pi }r_g \frac{U_m}{{E}_w} \text{ , }
\end{align}
where $U_m$ denotes the energy density of the large-scale magnetic field, $v$ is the particle velocity, and $r_g$ the gyro-radius of the particle.

As initial condition, we used a turbulence spectrum derived from the diffusion coefficient, as suggested by Galactic propagation modeling \citep{2011ApJ...729..106T}, 
\begin{align}
    D_0 &= 10^{28}\left(\frac{pc}{10\,\text{GeV}}\right)^{1/3}\left(\frac{B_0}{3\,\mu\text{G}}\right)^{-1/3} \text{ . }
\end{align}
However, the initial diffusion coefficient is reduced by a factor of ten on account of numerical constraints.

We use a growth-rate based on the resonant streaming instability \citep{Skilling.1975a, 1978MNRAS.182..147B},
\begin{align}
    \Gamma_g &= A\cdot\frac{v_\text{A}p^2v}{3E_\text{w}}\left|\frac{\partial N}{\partial r}\right| \text{ , }\label{eq:growth}
\end{align}
where $v_\text{A}$ is the Alfv\'en velocity. We introduced a linear scaling factor, $A$, to artificially enhance the amplification. We used $A=10$ unless stated otherwise in this paper to mimic the more efficient amplification due to the nonresonant streaming instability \citep{2000MNRAS.314...65L, 2004MNRAS.353..550B}. A value of 10 seems to be consistent with observations of historical SNRs \citep{2021A&A...654A.139B} and estimates of the growth-rates operating at the early stages of CR-acceleration \citep{2018MNRAS.479.4470M}. Furthermore, the particulars of the nonresonant amplification are beyond the present capabilities of our code. The nonresonant instability saturates by a modification of the bulk flow or a back-reaction of the thermal plasma to CR-streaming \citep{2009ApJ...694..626R, 2010ApJ...709.1148N,2017MNRAS.469.4985K}. In addition, the scattering efficiency of the nonresonant modes is quite unclear.

Since we exceed the growth rate of the resonant streaming instability \citep{1978MNRAS.182..147B} by a factor of ten, the turbulent field is amplified to $\delta B\gg B_0$. 
The timescale on which the turbulence in the precursor needs to be replenished is typically shorter than the time it takes for the resonant modes to saturate \citep{2016A&A...593A..20B, 2021ApJ...922....7I}, and so the peak magnetic field that we obtain is lower than the predicted saturation-level of $\delta B\approx \sqrt{v_\mathrm{sh}U_\mathrm{cr}/c} \geq B_0$  \citep{2000MNRAS.314...65L, 2001MNRAS.321..433B}. At the saturation amplitude $\sqrt{v_\mathrm{sh}U_\mathrm{cr}/c} $ for Bell modes throughout the cosmic-ray precursor one would not even get a single exponential growth cycle \citep{2021ApJ...921..121P}. Hence, the magnetic field in our simulation will still depend on the initial magnetic-field strength in the progenitor's wind.

Cascading is balancing the growth of the magnetic turbulence and hence the magnetic field level. This process is described as a diffusion process in wavenumber space, and the diffusion-coefficient is given by \citep{1990JGR....9514881Z, Schlickeiser.2002a}
\begin{align}
    D_\text{k} &= k^3v_\text{A}\sqrt{\frac{E_\text{w}}{2B_0^2}} \text{ . }
\end{align}
This phenomenological treatment will result in a Kolmogorov-like spectrum, if cascading is dominant. The cascading-rate will depend on the level of turbulence in two regimes, since $v_\text{A}\propto B_\text{tot}$
\begin{align}
    D_k \propto 
    \begin{cases}
        \sqrt{E_\text{w}} &\text{for $E_\text{w} \is{\ll} B_0^2/8\pi$}\\
        \is{E_\text{w}} &\text{for $E_\text{w} \is{\gg} B_0^2/8\pi$} \text{ . }\\ 
    \end{cases}\label{eq:Cascading}
\end{align}
When the turbulent field starts to dominate over the background field, the cascading rate depends more sensitively on the energy density of magnetic turbulence. 

\subsection{Hydrodynamics}\label{sec:Hydro}
The evolution of an SNR without CR feedback can be described with the standard gas-dynamical equations.
\begin{align}
\frac{\partial }{\partial t}\left( \begin{array}{c}
                                    \rho\\
                                    \textbf{m}\\
                                    E
                                   \end{array}
 \right) + \nabla\left( \begin{array}{c}
                   \rho\textbf{v}\\
                   \textbf{mv} + P\textbf{I}\\
                   (E+P)\textbf{v} 
                  \end{array}
 \right)^T &= \left(\begin{array}{c}
                    0\\
                    0\\
                    0
                   \end{array}
 \right)\\
 \frac{\rho\textbf{v}^2}{2}+\frac{P}{\gamma-1}  &= E \text{,}
\end{align}
where $\rho$ is the density of the thermal gas, $\textbf{v}$ the plasma velocity, $\textbf{m}=\textbf{v}\rho$ the momentum density, $P$ the thermal pressure of the gas and $E$ the total energy density of the ideal gas with $\gamma=5/3$. This system of equations is solved under the assumption of spherical symmetry in 1-D using the \textsc{PLUTO} code \citep{2007ApJS..170..228M}. It has to be noted, that radiative losses will play an important role in the early evolution of the remnant, especially in a very dense circumstellar medium (CSM). However, photons can not easily escape and a different equation of state has to be used to accurately describe regions in the far-downstream of the forward shock \citep{2019A&A...622A..73O}.

The results of the hydro simulation for the density, velocity, pressure, and temperature distributions are then mapped onto the spatial coordinate of the CR and turbulence grid, respectively. 
The shock, typically a few grid cells wide in the hydro-solution, needs to be resharpened in order to guarantee a realistically high acceleration rate from GeV to TeV energies. This procedure is repeated for each time step of the CR and turbulence grid. One of these time steps typically requires many time steps of the hydro solver. 

To improve the computational performance a dynamical re-gridding scheme was introduced, that dynamically moves the grid boundaries together with the shock structures. We use 16384 grid-cells in \textsc{PLUTO} that initially span the range between $8\cdot10^{12}$\,cm to $1.1\cdot10^{15}$\,cm. Over the course of 20 simulated years, the hydro-grid is remapped 20 and 7 times for the RSG and LBV-progenitor respectively.   

\subsubsection{Initial conditions for stellar wind and supernova}
The remnants are modeled as CC events following the initial conditions of \cite{1982ApJ...258..790C},
\begin{align}
 \rho(r) &= \begin{cases}
             \rho_{\mathrm c}, & r<r_{\mathrm c},\\
             \rho_{\mathrm c}\left(\frac{r}{r_{\mathrm c}}\right)^{-n}, & r_{\mathrm c}\leq r \leq R_{\mathrm{ej}},\\
             \frac{\dot{M}}{4\pi V_\mathrm{w} r^2}, & r \geq R_{\mathrm{ej}},\\
            \end{cases}
	\label{gasdyn}
\end{align}
The velocity of the ejecta is defined as
\begin{equation}
v_{\mathrm{ej}}(r) = \frac{r}{T_{\mathrm{SN}}},
\end{equation}
where $T_{\mathrm{SN}}$ is the initial time set for hydrodynamic simulations. We define the radius of the ejecta as a multiple of $r_{\mathrm c}$, $R_{\mathrm{ej}} = xr_{\mathrm c}$. Then, the initial conditions for simulations can be written as functions of the ejecta-mass $E_\text{ej}$ and the explosion energy $E_\text{ej}$
\begin{align}
  r_c &=
  \left[\frac{10}{3}\frac{E_{\mathrm{ej}}}{M_{\mathrm{ej}}} \left(\frac{n-5}{n-3}\right) \left(\frac{1- \frac{3}{n} x^{3-n}}{1-\frac{5}{n} x^{5-n}} \right) \right]^{1/2} T_{\mathrm{SN}}, \\
  \rho_c &= \frac{M_{\mathrm{ej}}}{4\pi r_c^3}\frac{3(n-3)}{n} {\left(1- \frac{3}{n} x^{3-n}\right)^{-1}}, \\
  v_{\mathrm c} &= \frac{r_{\mathrm{c}}}{T_{\mathrm{SN}}}. 
\end{align}
We used the SN parameters given in Table~\ref{tab:ProgenitorModels} and an explosion energy of $E_\text{ej}=10^{51}$\,erg for the different progenitor stars.

The stellar wind properties ($\dot{M}$ and $V_\mathrm{w}$) in Table~\ref{tab:ProgenitorModels} determine the CSM density that the SN ejecta interact with.
We assume in this work a steady mass-loss from the progenitor star, and so the density is monotonically decreasing with distance $r$ from the star according to Equation~\ref{gasdyn} for $r>R_\mathrm{ej}$.
The first two cases (LBV and RSG) in Table \ref{tab:ProgenitorModels} feature quite extreme values for $\dot{M}$, and in particular the winds of LBVs have no stable mass-loss at this rate \citep{2012ASSL..384..221V}.
Nevertheless these cases are useful for exploring the parameter space of CSM densities.
The last two cases have significantly lower mass-loss rates and correspond to more commonly observed wind properties.

The total mass of material in the CSM is important not only for the shock dynamics but also for absorption in the ambient medium, for instance of X-rays and radio waves (see section \ref{sec:Absorption}).
We assumed a total mass of $10M_\odot$ in the ambient medium around both progenitor stars for the LBV, RSG and LBV-moderate scenarios.
This means, that the RSG-scenario as a free wind-zone extending up to $\approx2$\,pc and the LBV to only a fraction of that.
For the RSG-moderate scenario we extend the wind to $\approx2$\,pc, with a total mass of only $\approx0.1\,\mathrm{M}_\odot$.

\begin{table*}
  \centering
  \caption{Parameters for the progenitor stars winds and initial remnant sizes \citep{2014MNRAS.440.1917D, 2021A&A...647A..99G, 2021ApJ...908...75B}.
  Columns 2-3 give the mass-loss rate (2) and wind velocity (3) (assumed constant) for each of the 4 cases, and colums 4-6 give the initial radius of the SN ejecta at the start of the simulation (4), the total ejecta mass of the SN (5) and the parameter $n$ determining the radial dependence of the ejecta (6) (see text for details).
  }
  \label{tab:ProgenitorModels}
  \begin{tabular}{c|c c c c c}
    Model & $\dot{M} [M_\odot/$yr] & $V_\mathrm{w}$ [km/s]  & $R_\text{ej}$ [cm] & $M_\text{ej}$ [$M_\odot$] & $n$ \\
    \hline
    LBV  & $10^{-2}$ & 100 & $1.2\cdot10^{14}$ & 10 & 10\\
    RSG  & $8.3\cdot10^{-5}$ & 15 & $6\cdot10^{13}$ & 3 & 9\\
    LBV moderate & $3\cdot10^{-4}$ & 100 & $1.2\cdot10^{14}$ & 10 & 10\\
    RSG moderate & $10^{-6}$ & 15 & $6\cdot10^{13}$ & 3 & 9\\
  \end{tabular}
\end{table*}

\subsection{Radiation and absorption processes}\label{sec:Absorption}
Due to the high densities, the material in and around the SN-explosion is not necessarily transparent to emission produced inside the SNR.
In order to calculate the observable radiation from such young objects the absorption for X-ray, radio and $\gamma$-ray radiation needs to be taken into account. 

We therefore attenuate the flux emitted from every part of the remnant by the optical depth produced by the matter that has to be transversed by the rays along the line-of sight according to 
\begin{align}
    F(x,E) &= F_0(x,E)\exp\left(-\tau(x,E)\right) \text{ , }
\end{align}
where $\tau$ is the optical depth and $F$ and $F_0$ are the attenuated and unattenuated photon-fluxes at a given location and energy respectively. The total flux is then obtained by summing the contributions from all parts of the remnant.  

\is{
\subsubsection{Non-thermal emission}
To calculate non-thermal emission from the SNR we consider synchrotron radiation in the radio and X-ray energy bands and proton interactions with subsequent pion decay in the gamma-ray energy band. Gamma-ray emission produced through the inverse Compton scattering in considered scenarios appears to be negligible comparing to that from hadronic interactions and therefore was ignored. 
}

\subsubsection{Thermal X-ray emission} \label{sec:Xray}
We model the thermal continuum X-ray emission using the expressions derived by \cite{1999A&A...344..295H}. We follow the results of \cite{2014MNRAS.440.1917D} and consider only the emission from the forward shock of the remnant. It is suggested that a dense, low-temperature shell forms around the reverse shock at the early stages of the SNR evolution and high-density environments that are considered here. We thus exclude any thermal emission from regions inside of $r<0.85\cdot R_\text{sh}$. We account for the absorption of the X-ray emission outside the remnants by the dense shells surrounding them. The optical depth is calculated according to 
\begin{align}
    \tau &= \int_x^{\text{R}_\text{Shell}}n_H(x^\prime)\sigma_E\text{d}x^\prime \text{ , }
\end{align}
where $n_H$ is the hydrogen density at a given location and $\sigma_E$ the cross section \citep{2019MNRAS.486.1094M,1983ApJ...270..119M}
\begin{align}
    \sigma_E &= 2.27\cdot10^{-22}E_{keV}^{-2.485}\,\text{cm}^2 \text{ , }\label{eq:SigmaX}
\end{align}
where $E_{keV}$ is the photon-energy in keV. Equation (\ref{eq:SigmaX}) approximates losses due to absorption by ions of heavy elements in a gas with approximately solar abundances. This is of course an over-simplification of the process since ionization plays an important role e.g. by the SN flash, which cannot be accounted for in our model. However, it was suggested in the literature, that up to $2M_\odot$ of material can be ionized by the SN emission \citep{1996ApJ...464..924L}. During the absorption calculation, we estimate the size of this ionized region and take only material outside the ionized shell into account as attenuating the X-ray emission. Still, we might over-estimate the absorption of thermal X-rays. However, the scope of this calculation is to get an idea of the thermal X-ray emission and allow comparisons to other wavelengths, especially with respect to the timing of the emission.

\subsubsection{Free-free absorption}
Free-free absorption (FFA) is thought to dominate for SNRs in dense circumstellar environments \citep[and references therein]{2021ApJ...908...75B} at radio frequencies. \cite{1967ApJ...147..471M} found that the absorption coefficient, $\kappa$, for FFA can be expressed as
\begin{align}
    \kappa &= 3.3\cdot10^{-7}\left(\frac{n_e}{\text{cm}^{-3}}\right)^2\left(\frac{T_e}{10^4 \text{K}}\right)^{-1.35}\left(\frac{\nu}{\text{GHz}}\right)^{-2.1}\text{pc}^{-1},
\end{align}
where $n_e$ denotes the electron density, $T_e$ the electron temperature and $\nu$ the frequency of the emitted photon. The optical depth follows from integrating $\kappa$ over the line-of-sight,
\begin{equation}
    \tau_\mathrm{ff} = \int_\mathrm{LoS}\kappa dx
\end{equation}

We calculate the synchrotron emission from our non-thermal population of electrons and the total magnetic field inside the remnant and attenuate the respective flux according to the material that has to be traversed. 

\subsubsection{$\gamma\gamma$-absorption}
\label{gg_absorption}
The $\gamma$-ray photons emitted by the shock-accelerated CRs might undergo pair-production in the intense photon fields of the SN photosphere. \cite{2009A&A...499..191T} showed that this process can significantly attenuate the VHE $\gamma$-ray flux for the first year after the explosion. While this first work relied on the assumption of an isotropic photon-field, \cite{2020MNRAS.494.2760C} calculated the attenuation by anisotropic $\gamma\gamma$ interactions also accounting for temporal effects connected to the variation of the photosphere with time.
They showed that the VHE-flux is lowered most strongly during the first tens to hundreds of days depending on the energy band and hence less strongly than predicted by isotropic interactions.
While both mentioned works focused on SN 1993J, the brightest radio SN on record, where a detailed monitoring has taken place across many wavebands, we consider here more generic cases. Hence, we ignore the temporal evolution of the photosphere and corresponding light-travel effects that can play a role for the VHE-attenuation prior to the 10-days time-scale. However, we still account for the anisotropic nature of $\gamma\gamma$ interactions and for the finite size of the photosphere. 
Furthermore, unlike \citet{2020MNRAS.494.2760C} where the gamma-ray emission is restricted to a thin shell and the optical depth is calculated for the whole remnant, we do account for the spatial distribution of emitting particles by calculating the optical depth for every spatial location within the radius of $r_\mathrm{max} = 1.2R_\mathrm{sh}(t)$, where $R_\mathrm{sh}(t)$ is the shock radius at a certain age. (see Appendix for details).

For calculation of the $\gamma\gamma$ absorption we roughly follow the approach introduced for binary systems by \citet{2006A&A...451....9D} and farther used and extended in \citet{2017ApJ...837..175S}. The \gaga\ optical depth seen by a $\gamma$-ray photon of energy $E_\gamma = \epsilon_\gamma  m_\mathrm{e} c^2$ traveling over the distance $l$ is given by \citep{1967PhRv..155.1404G}
\begin{equation}
  \tau_{\gamma\gamma} = \int_0^l\,{\rm d}l \int_{4\pi}  {\rm d}\Omega \, \,(1-\mu) \int_{\frac{2}{\epsilon_\gamma (1-\mu)}}^\infty {\rm d} \epsilon \, n_\mathrm{ph}(\epsilon,\Omega) \sigma_{\gamma\gamma}(\epsilon,\epsilon_\gamma,\mu)
\label{eqn:tau_gamma}
\end{equation}
where $\mu = \cos \theta$ with $\theta$ being an interaction angle between the $\gamma$-ray photon and low-energy target photon, ${\rm d}\Omega = \sin{\theta^\prime}d\theta^\prime d\phi^\prime$ is the solid angle element in the spherical coordinate system centered at the $\gamma$-ray photon with zenith determined by the direction from the centre of the photosphere (see Appendix~\ref{ggCalculation}), 
$\epsilon$ is the energy of the target photon normalized to the electron rest mass ($\epsilon = h\nu/(m_\mathrm{e} c^2)$), and $n_{\rm ph}(\epsilon,\Omega)$ is the number density per unit solid angle of the low-energy target photons. 
The \gaga\ cross section $\sigma_{\gamma\gamma}$ is given by \citep{1976tper.book.....J},
\begin{equation}
 \sigma_{\gamma\gamma}(\beta) = \frac{3}{16} \sigma_{\rm T} ( 1- \beta^2) \left[ (3-\beta^4) \ln \left(\frac{1+\beta}{1-\beta} \right) -2\beta (2-\beta^2)\right],
\end{equation}
where 
\begin{equation}
 \beta = \sqrt{1-\frac{2}{\epsilon \epsilon_\gamma (1-\mu)}},
\end{equation}
and $\sigma_{\rm T}$ is the Thomson cross-section.

\begin{figure*}
    \centering
    \includegraphics[width=0.99\textwidth]{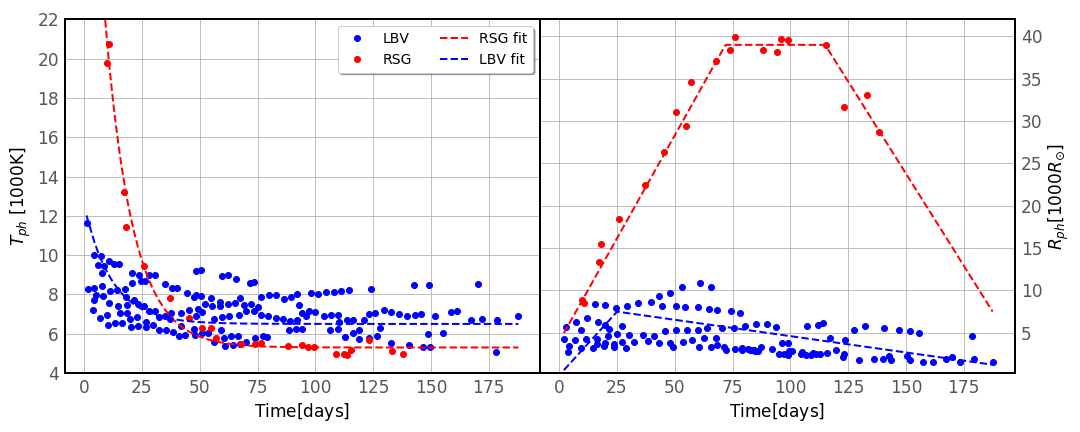}
    \caption{Photosphere temperature (left) and photosphere radius (right) for Type-IIn/LBV (blue) and Type-IIP/RSG SNe (red). The data (points) is taken from \protect\cite{2020A&A...638A..92T} and \protect\cite{10.1093/mnras/sty1634} respectively.}
    \label{fig:Photospheres}
\end{figure*}

We extracted the photospheric properties for Type-IIn and Type-IIP explosions from observations of \cite{2020A&A...638A..92T} and \cite{10.1093/mnras/sty1634} respectively.
We fitted the photospheric radius with piecewise linear functions and described the cooling of the photosphere by an exponential function. The fit-functions are shown alongside the observational data in Figure \ref{fig:Photospheres}. 
The photospheric target photon field is assumed to follow a blackbody distribution where the photon density is therefore given by
\begin{equation}
    n_\mathrm{ph}(\nu,\Omega) = \frac{2\nu^2}{c^3}\frac{1}{e^\frac{h\nu}{kT_\mathrm{ps}}-1}.
\end{equation}

We used the information on the photospheric properties to calculate the opacity for photons emitted in different parts of the remnant and obtain the attenuated $\gamma$-ray spectra. 


\subsubsection{EBL-absorption}
For extragalactic sources another photon field which could significantly contribute to the gamma-gamma absorption process discussed in the section \ref{gg_absorption} is the extragalactic background light \citep[EBL; see e.g.][]{1966PhRvL..16..252G}.
Therefore, we take it into account while studying the detectability of distant Type-IIP and Type-IIn explosions.
For the distribution of the EBL we use the prescription suggested by \cite{2010ApJ...712..238F} and calculate the $\gamma\gamma$ optical depth 
of the universe, $\tau_{\gamma\gamma}^{\mathrm{EBL}}$, for gamma-ray photons with energies corresponding to the upper and lower energy-threshold of various gamma-ray observatories depending on the distance $d$ using Astropy \citep{astropy:2018}. We then average the values for both energies and apply the result as a correction factor of $e^{-\tau_{\gamma\gamma}^{\mathrm{EBL}}}$ to the gamma-ray flux.

However, we find that the absorption is still negligible for the relevant distances and energies. The highest impact is found for observations with \emph{LHAASO}, where at most $\approx10\,$\% of the gamma-ray flux is absorbed. A summary of the relevant parameters is shown in Table \ref{tab:EBL}

\begin{table}
  \centering
  \caption{The impact of EBL-absorption on the observations of distant SNRs with different gamma-ray observatories.}
  \label{tab:EBL}
  \begin{tabular}{c|c c c c }
    Instrument & $E$ [TeV] & $d_\text{max}$ [Mpc] &  $\exp(-\tau)$ \\
    \hline
    \emph{Fermi}  & 0.001 - 0.3 & 3 & >0.99 \\
    \emph{H.E.S.S.}  & 1 - 10 & 3 & 0.986 \\
    \emph{HAWC} & 2 - 20 & 1.5 & 0.988 \\
    \emph{CTA-south} & 1 - 20 & 12 & 0.92 \\
    \emph{LHAASO} & 40 - 300 & 6 & 0.87 \\
  \end{tabular}
\end{table}
\section{Results}
\label{sec:results}
\begin{figure*}
    \centering
    \includegraphics[width=0.99\textwidth]{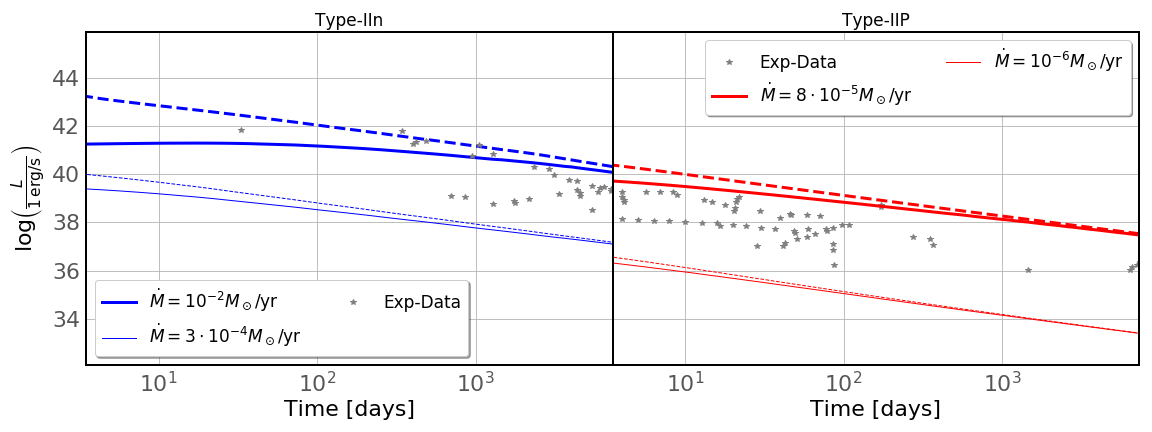}
    \caption{Comparison of the thermal X-ray emission, where the unabsorbed flux is shown in dashed lines and the flux including absorption in solid lines. The thick lines represent the cases of higher mass-loss of the progenitor, while the thin lines represent the moderate scenarios from table \ref{tab:ProgenitorModels}. The experimental data is taken from a compilation of X-ray measurements from many SNe by \citet{2014MNRAS.440.1917D}.}
    \label{fig:ThermalX}
\end{figure*}

We followed the evolution of the remnants for 20 years. In order to verify the hydrodynamic evolution of our remnants, we calculated the thermal X-ray continuum emission as described in section \ref{sec:Xray}. Figure \ref{fig:ThermalX} shows the evolution of the X-ray emission compared to measurements from Type-IIn and Type-IIP SNe respectively.
It can be seen that the absorption is negligible for Type-IIP explosions even in the case of high mass-loss. This is consistent with estimates of the column-density for known Type-IIP SNe \citep{2007MNRAS.381..280M}. The X-ray flux that we obtain is consistent with the data collected by \citet{2014MNRAS.440.1917D}. The Type-IIP progenitors of the SNe shown in Figure \ref{fig:ThermalX} have estimated mass-loss rates of $[0.3-10]\cdot10^{-6}\,M_\odot$\,yr$^{-1}$, right between our moderate and high mass-loss scenario.  

Figure \ref{fig:ThermalX} indicates that the absorption is more important in the case of Type-IIn explosions. Especially in the beginning, the flux is attenuated by $\approx2$ orders of magnitude for the highest mass-loss rate. We are simplifying the situation for LBVs by assuming a steady mass-loss rate and hence a smooth density distribution around the progenitor star. LBVs are known for episodes of extensive mass-loss with $\dot{M}$ reaching values of up to $1\, M_\odot$\,yr$^{-1}$ \citep{2012ASSL..384..221V}.
This means a comparison with observations is not straightforward, but Figure~\ref{fig:ThermalX} indicates that our assumed mass-loss rates are in reasonable agreement with the observed X-ray luminosities.

\subsection{Maximum particle energies}
\begin{figure}
    \centering
    \includegraphics[width=0.49\textwidth]{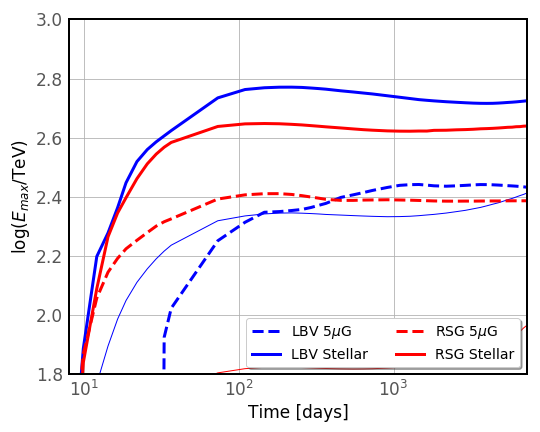}
    \caption{Maximum energy of protons for a LBV progenitor (blue) and a RSG progenitor (red).
    Solid lines indicate a $1/r$ magnetic field derived from the progenitors surface field and dashed lines indicate models with a constant $5\,\mu$G upstream field. The thick lines represent the high mass-loss cases from Table \ref{tab:ProgenitorModels} and the thin lines the moderate cases. }
    \label{fig:MaxE}
\end{figure}

At each time step, the simulated proton spectrum is fitted with
\begin{align}
    N_\text{fit} = p^{-s}\exp\left(-\left(\frac{p}{\is{p_\mathrm{max}}}\right)^a\right)\text{.}\label{eq_cutoff}
\end{align}
The time evolution of the fitting parameter, $E_\text{max}=\sqrt{(p_\mathrm{max} c)^2 +(m_0 c^2)^2}$, for our four different configurations is shown in Figure \ref{fig:MaxE}.
It is evident that a higher ambient density is increasing the maximum energy (LBV vs. RSG) since a denser CSM means that more particles get injected into the acceleration process as CRs. A higher CR density translates into more driving of turbulence by the CR gradient in equation (\ref{eq:growth}). In this early stage of the evolution, the shock speed is only weakly dependent on the ambient density. The density dependence is stronger in the Sedov stage where the effect of potentially more injected CRs gets almost completely canceled by a correspondingly lower shock speed.
In our simulations both progenitor scenarios saturate at roughly the same maximum energy.
The 2$\times$ higher shock velocity in the RSG scenario compensates for the approximately 5 times lower level of the amplified field, since the acceleration efficiency depends on $v_\text{sh}^2$.

A slight increase of $E_\text{max}$ after $\approx3\,$years for some of the cases is caused by a change of the cutoff-shape. During the initial stage of the acceleration the cutoff is not strictly exponential as the fitting parameter $a$ decreases from $a\approx10$ initially, to $a\approx3$ towards the end of the simulations\footnote{The effect of the changing shape of the particle cutoff is described in more detail in section 5.1.1 and 5.1.2 in \cite{2016A&A...593A..20B}}. However, equation (\ref{eq_cutoff}) is still just approximating the shape of the cutoff, so that a part of the change in the cutoff-shape gets absorbed into the value of $E_\text{max}$ which is increasing slightly as a result.  

At the same time, the higher ambient magnetic field translates into a higher particle energy. In general, the higher field means there is a higher growth of turbulence due to the higher Alfv\'en speed. However, once the level of turbulence exceeds $\delta B/B_0=1$, there is a significant contribution from the magnetic turbulence to the background field and as a consequence, the scaling of the cascading term for the magnetic turbulence changes as illustrated in Eq. \ref{eq:Cascading}.
In the strong turbulence regime the cascading \is{term} increases faster \is{($\propto B_\text{tot}^{4}$)} than the growth \is{term} ($\propto B_\text{tot}$). This limits the maximum field to values below the values predicted by works which only consider the saturation level of the non-resonant instability as the limiting factor for the amplification \citep{2014ApJ...789..137B, 2018MNRAS.479.4470M}. In addition to the neglect of the cascading, the studies of \cite{2014ApJ...789..137B} and \cite{2018MNRAS.479.4470M} assume a steady-state configuration. In reality, the time to build turbulence in the precursor is limited by the timescale $v_\text{sh}/D$. This limited time to grow turbulence further reduces the maximum energy that particles can achieve \citep{2016A&A...593A..20B}. \cite{2021ApJ...922....7I} showed that this is even true without the effect of cascading and that the non-resonant mode is typically not entering the saturation regime.

As a consequence, the maximum achievable energies are well below the PeV energies needed for the sources of Galactic CRs but can reach $600\,$TeV when a strong ambient stellar field is present. This is below the values obtained by \cite{2021ApJ...922....7I}, where despite the time-limited growth, particle energies above $1\,$PeV where obtained for configurations with mass-loss rates of $\dot{M}=10^{-3}\,M_\odot$\,yr$^{-1}$. 
This difference can not be accounted for by a difference in the injection-parameters since the values spanned by \cite{2021ApJ...922....7I} are in the range of $\eta_i=[2-6]\cdot10^{-6}$, compatible with $\eta_i=3.6\cdot10^{-6}$ in our simulations. More moderate mass-loss rates, closer to the population-average for LBV and RSG progenitor stars, result in even lower maximum energies of $\approx 230\,$TeV and $\approx 70\,$TeV for LBV and RSG progenitors respectively.

The stellar fields assumed in this work are reasonable in comparison with measurements from RSGs \citep{2017A&A...603A.129T} and energetic arguments for LBVs \citep{WhiBreKon20} but cannot be dramatically increased.
Mass-loss rates span the typical range up to extreme large values for our high-mass-loss cases. Our findings therefore support the suggestion by \cite{2020APh...12302492C} that only exceptionally powerful (and unusual) events are able to reach energies needed for the knee of Galactic CRs, where the explosion energy is above the canonical $10^{51}\,$erg and the ejecta-mass is low, to allow for faster propagating shocks. However, the ARGO-YBJ collaboration reported the position of the H\&He-knee to be around $700\,$TeV, which is more in line with our findings \citep{2015PhRvD..92i2005B}.


\subsection{Gamma-ray emission}
\label{sec:gammarays}
\begin{figure*}
    \centering
    \includegraphics[width=0.95\textwidth]{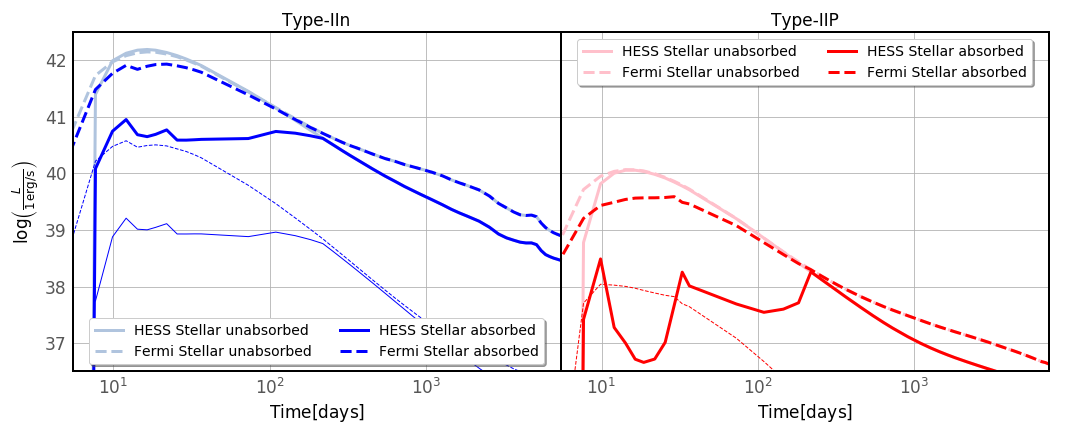}
    \caption{Gamma-ray luminosities in the \textit{Fermi-LAT} energy range (dashed) and \textit{H.E.S.S.} energy range (solid) for the stellar ambient field accounting for $\gamma\gamma$-absorption (strong colors) and without absorption (pastel colors). The thick lines represent the high mass-loss cases from Table \ref{tab:ProgenitorModels} and the thin lines the moderate cases. Note, middle peaks that can be seen in predicted absorbed H.E.S.S. light curved is the numeric artifact (see the main text and Appendix for details).}
    \label{fig:GammaL}
\end{figure*}

We derived the $\gamma$-ray luminosity in the 1-10\,TeV (hereafter \textit{H.E.S.S.}) and 1-300\,GeV (hereafter \textit{Fermi}) energy bands.
Figure \ref{fig:GammaL} illustrates the evolution of the $\gamma$-ray luminosity for our models.
We start the acceleration of CRs three days after the explosion, based on numeric constraints. Despite the high growth-rate of turbulence in the stellar-field case, it needs about 10-15 days for the intrinsic $\gamma$-ray luminosity in the \textit{H.E.S.S.} band to reach its peak value. The intrinsic luminosity peaks marginally earlier in the \textit{Fermi} band due to the time it needs to build up turbulence resonant with particles of the highest energies. It should also be stressed here that the middle spike appearing in the absorbed \textit{H.E.S.S.} light curve at 25 days for the Type-IIn progenitor and at 30 days for the Type-IIP progenitor is a numeric artifact arising from the interpolation between $\tau_{\gamma\gamma}$ matrices calculated for different times (see Appendix for details).

For both progenitor cases, the intrinsic $\gamma$-ray luminosity is significantly reduced by $\gamma\gamma$-absorption due to the SN photosphere. The effect is, however, much more pronounced in the \textit{H.E.S.S.} energy band, because for the considered parameters of the photosphere the optical depth is strongest for gamma-rays with energies above 1~TeV. While the \textit{Fermi} peak luminosity is reduced only by a factor of 2 at most, the \textit{H.E.S.S.} peak luminosity is reduced at least by an order of magnitude. Only at very early times (up to $\sim$15 days) after the explosion of the Type-IIP/RSG SN the absorption is most efficient for sub-TeV gamma-rays due to the high temperature of the photosphere in this case. At $\sim$10 days the peak absorption starts shifting to the \textit{H.E.S.S.} energy band which is clearly visible from the light curve with the flux attenuation up to 3 orders of magnitude that happens shortly after (see right panel of Fig.~\ref{fig:GammaL}; strong solid red curve). For both types of progenitors the peak of the intrinsic luminosity at about 15 days after explosion falls in the time interval where the absorption is the highest between $\sim10$ and $\sim220$ days which in turn correspond to times when expected absorbed luminosity is at its maximum.
The peak of predicted absorbed luminosity reaches $6.3\cdot10^{40}\,$erg\,s$^{-1}$ and $7.2\cdot10^{41}\,$erg\,s$^{-1}$ for the \textit{H.E.S.S.} and \textit{Fermi} band, respectively, for a Type-IIn progenitor, and $2\cdot10^{38}\,$erg\,s$^{-1}$ and $3.5\cdot10^{39}\,$erg\,s$^{-1}$ for a Type-IIP progenitor. The intrinsic luminosity is decreasing roughly as $1/t$ after its peak in all the models.



The prediction of the $\gamma$-ray luminosity in the \textit{H.E.S.S.}\ band is in line with current observational upper limits of $\leq1.5\cdot10^{39}\,$erg\,s$^{-1}$ for nearby Type-IIP events \citep{2019A&A...626A..57H}.
Figure \ref{fig:GammaL} also shows the absorbed gamma-ray luminosity for models with more modest mass-loss rates. In these cases, the luminosity reaches $\approx10^{39}\,$erg\,s$^{-1}$ for a Type-IIn and $\approx10^{36}\,$erg\,s$^{-1}$ for a Type-IIP explosion.
Since our predicted luminosity is roughly a factor of 100 below the current observational limit, the detection prospects are dim unless a exceptional nearby SN occurs or happens in an extremely dense CSM. Further, the prospects for any detection of a Type-IIP event are small before the SN photosphere stops increasing in luminosity, around $40\,$ days post-explosion.
We evaluate the detection prospects for different current and next-generation instruments in the next section.

\subsubsection{Fermi-LAT detectability}
\begin{figure*}
    \centering
    \includegraphics[width=0.91\textwidth]{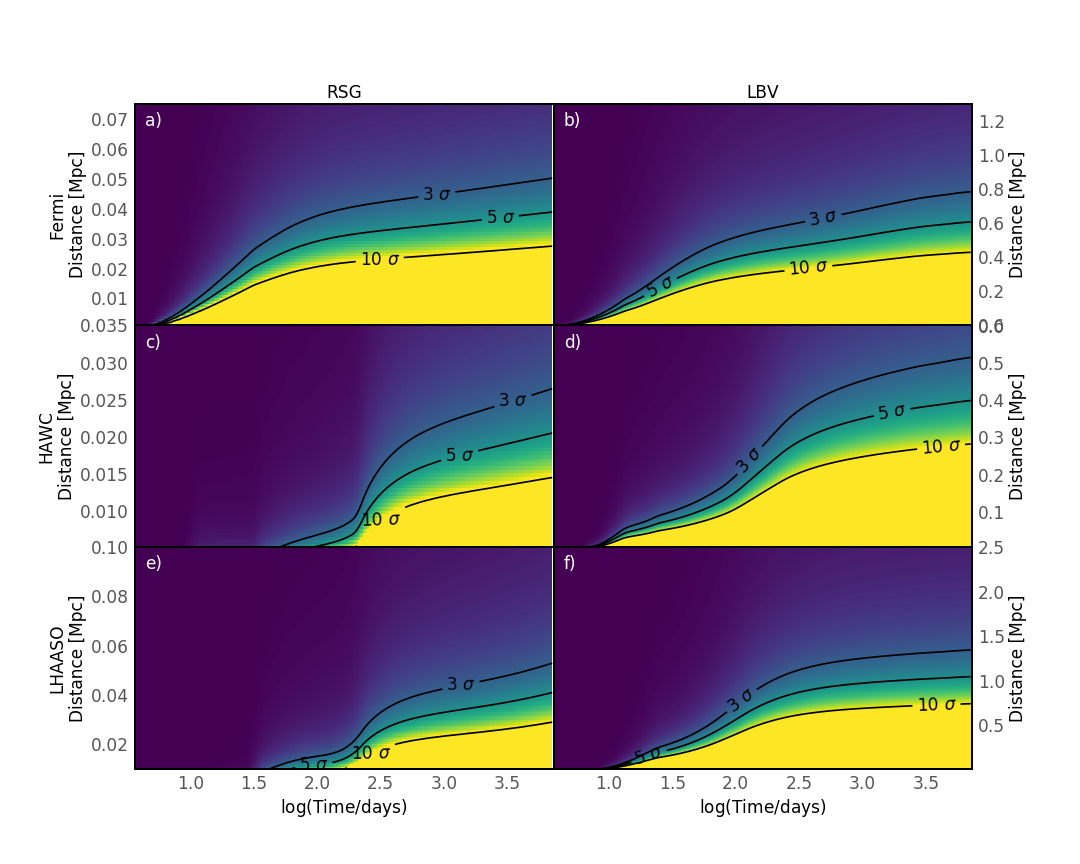}
    \caption{Detectability of nearby SNe by the survey-like experiments \emph{Fermi-LAT} (1-300GeV), \emph{HAWC} (2-20TeV) and \emph{LHAASO} (40-300TeV). Shown is the culminated significance for a SNR observed since explosion. The linear colorscale spans from $0$ to $10\sigma$.}
    \label{fig:Survey}
    \includegraphics[width=0.91\textwidth]{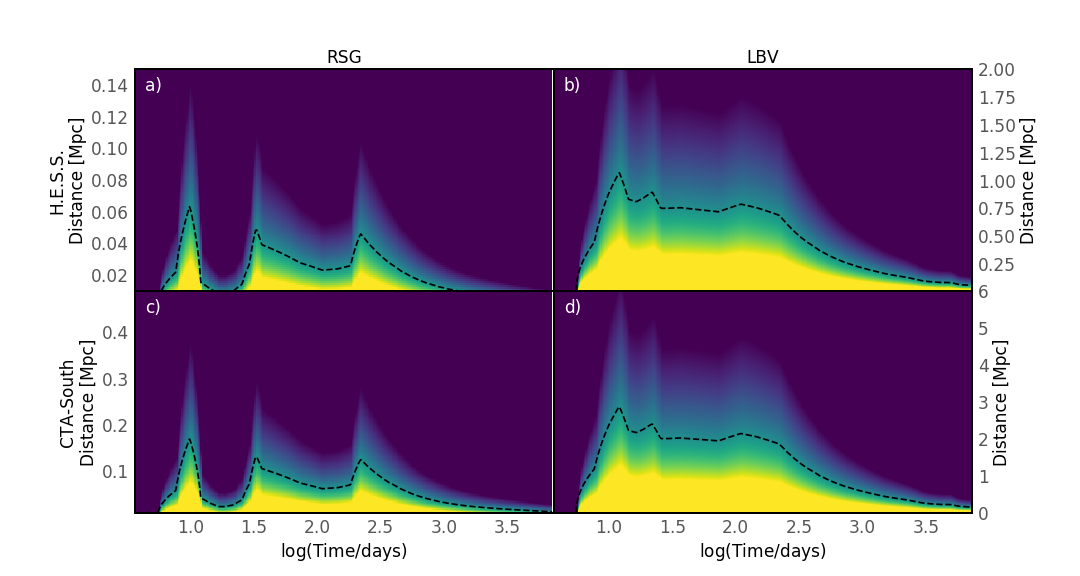}
    \caption{Detectability of nearby SNe by the IACTs \emph{H.E.S.S.}\ and \emph{CTA-South}. The logarithmic color scale shows flux as a function of time post-explosion and of distance to the SN, where lighter colors denote a higher flux. The dashed lines indicate the flux needed for a $5\sigma$ detection within 50hrs for both instruments. The color-scale spans from 1/5th to five-times the detection-flux in all cases. Note, middle peaks that can be seen in all four panels is the numeric artifact (see the main text and Appendix for details).}
    \label{fig:IACTs}
\end{figure*}
We investigate the prospects for a detection of a $\gamma$-ray signal in the \textit{Fermi}-band. Unlike the Imaging Air Cherenkov Telescopes (IACTs), \textit{Fermi-LAT} is constantly monitoring the sky and thus allows a retrospective study of potentially interesting objects.

Figure \ref{fig:Survey}, panels a) and b) show the integrated sensitivity that \emph{Fermi-LAT}\footnote{Obtained from https://fermi.gsfc.nasa.gov/ssc/data/analysis \mbox{/documentation/Cicerone/Cicerone\_LAT\_IRFs/LAT\_sensitivity.html}} would reach when observing the SNR since the time of explosion accounting for $\gamma\gamma$-absorption in the energy-range between $1-300\,$GeV. We calculated the total number of photons that \emph{Fermi-LAT} needs to collect for a detection of the source. However, we are not able to account for the impact of background-photons, hence the predicted increase of the significance later than $\approx200\,$days is likely overestimated.

Our prediction is that all supernovae expanding into wind density-profiles are already fading significantly after one month. As a result, the fast initial rise of the significance slows down with the onset of the fading. After this, the culminative significance continues to rise more slowly. The maximal distance up to which a \emph{Fermi} signal can be expected within the first $\approx100\,$days is $\approx600\,$kpc for Type-IIn explosions and $\approx30\,$kpc for Type-IIP explosions.

So far, there is no good candidate for a nearby Type-IIn or Type-IIP SN that should have been detected by \emph{Fermi-LAT}. The situation is complicated by the fact that many SNe occur in star-forming or star-burst galaxies that might already have significant background $\gamma$-ray emission. Thus, a variation of flux imposed by the SN is potentially harder to detect than an isolated, background-free, point source, which is assumed here. 



\subsubsection{VHE-detectability} 
The current-generation IACTs, where we picked \emph{H.E.S.S.}~as a representative, have the advantage that sensitivity can be accumulated over much shorter time periods than with \emph{Fermi-LAT} or ground-based water-Cherenkov detectors like \emph{HAWC} and \emph{LHAASO}.
As a result, the SNRs can easily be observed during the time of their peak-flux.
Figure \ref{fig:IACTs}, panel a) and b) illustrate the expected flux\footnote{Note, middle peaks in both figures are numeric artifacts as explained in Section \ref{sec:gammarays}} from the SNRs compared to the 50h observation sensitivity of H.E.S.S.~\citep{2016JPhCS.718e2043V}. The maximum distance up to which a nearby Type-IIn SNR could be detected is $\approx1.0\,$Mpc.
Type-IIP could only be detected within $60\,$kpc, limiting potential host-galaxies to the Milky Way and its nearby satellite galaxies including the Magellanic clouds. However, there is no SN since SN1987A recorded close enough to be detected by H.E.S.S. and the other IACTs.

HAWC shows a similar differential sensitivity as the IACTs for a $F\propto E^{-2}$ spectrum, albeit at higher energies \citep{2016JPhCS.718e2043V}. Hence, as the emission-spectrum is extending well beyond the most-sensitive energy-range for HAWC, the detection horizon is comparable to the one of current-generation IACTs and no detection should be expected so far (Fig.~\ref{fig:Survey}, panel c) and d)).

The prospects are somewhat better for the next-generation facility \emph{CTA-south} (Fig.~\ref{fig:IACTs}, panel c) and d)) \citep{2016JPhCS.718e2043V}. Here, the detection horizon is $3\,$Mpc and $0.2\,$Mpc for a Type-IIn and Type-IIP SNRs respectively. While Type-IIP explosion are still not expected to be detectable beyond the Milky Way and its satellite galaxies, the larger volume in which a Type-IIn could be detected makes a detection more likely. However, the rate of all Type-II SN at low redshifts is $3\cdot10^{-4}\,$yr$^{-1}$Mpc$^{-3}$, bringing the rate for Type-II SN that are close enough to one every $\approx80\,$yrs \citep{2012A&A...545A..96M}. Recent estimates for the local group yield a value of one SN every $\approx60\,$yrs \citep{2021NewA...8301498R}.

\subsubsection{UHE-detectability}
\emph{LHAASO} recently opened the window to ultra-high-energy (UHE) $\gamma$-ray astronomy \citep{2021ChPhC..45b5002A}.
We investigated the capability of \emph{LHAASO} to discover $\gamma$-rays of nearby SNRs similarly to the method for \emph{Fermi-LAT} and \emph{HAWC} (\ref{fig:Survey}, panel e) and f)) \citep{2016JPhCS.718e2043V}. It turns out that the detection threshold is $1\,$Mpc and $0.04\,$Mpc for a Type-IIn and Type-IIP SNRs respectively, giving it similar capabilities to \emph{CTA South}. However, the differential sensitivity of \emph{LHAASO} significantly increases beyond $100\,$TeV of $\gamma$-ray energy. In our models, the $\gamma$-ray spectra cut-off around $75\,$TeV, thus not benefiting from this high-sensitivity region. It can be expected, however, that \emph{LHAASO} is very sensitive to any SNR that accelerates CRs to higher energies, hence exceptional objects that expand in highly magnetized ambient media or feature a higher explosion energy, like superluminous SNe.  

\subsubsection{Comparison to other models}
The acceleration of CRs in a dense CSM has been modeled before. However, the luminosities that were obtained earlier are considerably larger than our estimates. \cite{2016APh....78...28Z} considered a non-linear acceleration model for Type-IIn explosion with parameters comparable to our setup albeit a factor of 10 higher explosion energy. Given the young age and hence the large shock ram-pressure, they needed to inject a higher fraction of thermal particles into the acceleration process to modify the shock structure and consequently obtained a unabsorbed gamma-ray luminosity that is about 5 orders of magnitude above our results. The phenomenological model by \citep{2014MNRAS.440.2528M} yields a similarly high unabsorbed luminosity with a resulting detection horizon of up-to $\approx200\,$Mpc for Type-IIn explosion in dense environments. 

Taking into account the \gaga-absorption further reduces the detection prospects. \cite{2009A&A...499..191T} calculated the gamma-ray luminosity and considered isotropic \gaga-absorption in the initial stages of the remnant's evolution. However, they still assume that already a high fraction of the explosion energy was being converted to CRs yielding an unattenuated gamma-ray flux about 5 times higher than we obtain. 

\cite{2020MNRAS.494.2760C, 2022MNRAS.511.3321C} accounted for the anisotropic nature of the \gaga-absorption in the case of SN 1993J and for Type-IIP explosions. This leads to a less-absorbed gamma-ray flux compared to the absorption-model of \cite{2009A&A...499..191T}. However, they still obtain a gamma-ray flux that is a factor of a few above our prediction.

The major difference between the models discussed here can be attributed to the fraction of thermal particles that get injected into the acceleration process as CRs and hence the fraction of the explosion energy that is converted into CRs, $\epsilon_\text{CR}/E_\text{ej}$, at the early stages of the remnant's evolution.
The values range from up to $30\%$ \citep{2014MNRAS.440.2528M}, $\approx10\%$ \citep{2009A&A...499..191T, 2020MNRAS.494.2760C, 2022MNRAS.511.3321C}, a few percent in a superluminous event \citep{2016APh....78...28Z} to $\approx0.1\%$ in our case. 
It needs to be mentioned that, in contrast to the other works, we do not use the energy-fraction of the CRs as an input variable to our model but obtain the values as a result from injecting a constant fraction of the thermal particles as CRs. We use an injection value that is consistent with observations of historical SNRs and yields an energy fraction of $\approx10\%$ of the explosion energy in CRs at the end of the Sedov-Taylor phase \citep{2020A&A...634A..59B, 2021A&A...654A.139B}. Models with high values for $\epsilon_\text{CR}/E_\text{ej}$ have difficulties in explaining the late-time emission from SNRs, as in their case basically no CRs are allowed to be accelerated after a few years on account of the limited energy-budget of the SNR and a fraction of $\epsilon_\text{CR}/E_\text{ej}\approx0.1$ that is sufficient to explain the energy-budget of Galactic CRs through SNRs.

\subsection{Radio emission}
\label{sec:radio}
\begin{figure}
    \centering
    \includegraphics[width=0.49\textwidth]{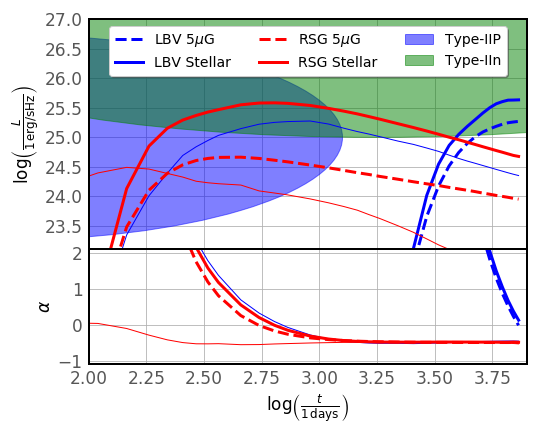}
    \label{fig:Radio}
    \caption{\textbf{Top panel:}Absorbed radio luminosity for a Type-IIn (blue) and Type-IIP (red) supernova progenitor. The solid lines indicate the stellar field scenario and the dashed lines the constant field scenario. The blue and green areas indicate the $1\sigma$ uncertainty region for the rise-time and peak radio-luminosity for Type-IIP and Type-IIn SN respectively. The thick lines represent the high mass-loss cases from Table \ref{tab:ProgenitorModels} and the thin lines the moderate cases. \newline
    \textbf{Bottom panel:} Radio spectral index $\alpha$ of the absorbed radio flux at 5\,GHz.}
    \label{fig:RadioIndex}
\end{figure}

We calculated the radio emission based on the electron distribution and the magnetic field, including the self-amplified component. Figure \ref{fig:RadioIndex} shows the radio luminosity at $8\,$GHz, including the effects of free-free absorption in the CSM, and the evolution of the radio spectral index.

A comparison to the population study of radio SNe from \cite{2021ApJ...908...75B} shows that our predicted peak radio luminosity is well within the observed range for both progenitor types. 
Both progenitors peak at about the same luminosity of $L_\text{Radio}\approx3\cdot10^{25}\,$erg\,s$^{-1}$\,Hz$^{-1}$ in our simulation, whereas the observed Type-IIn luminosity is a factor of 100 higher.
However, the stronger FFA absorption in the LBV case delays the rise of the radio emission to later times.
The intrinsic luminosities are considerably higher, especially in the LBV case, where an unabsorbed $L_\text{Radio}\approx3\cdot10^{26}\,$erg\,s$^{-1}$\,Hz$^{-1}$ is reached after about 1000 days post-explosion.
Since we are considering extreme cases of mass-loss, the fact that our radio rise-time is higher than that of the average sample of Type-IIP and Type-IIn SNe is not surprising.
On the other hand, the peak luminosities are within the observed range but appear low given the extreme mass-loss rates considered here.
Especially the peak-luminosity for Type-IIn explosions appears to be somewhat below the population average.
This can be attributed to either a too high mass in the absorbing CSM or the absence of further magnetic-field amplification in the post-shock region by plasma instabilities \citep[and references therein]{2014NIMPA.742..169F} in our model. 

In contrast, the cases for moderate mass-loss peak earlier for Type-IIn and Type-IIP explosion, in both cases in good agreement with the population average of the rise time.
The peak luminosity of the moderate-RSG progenitor reaches $L_\text{Radio}\approx3\cdot10^{24}\,$erg\,s$^{-1}$\,Hz$^{-1}$, close to the population-average of $L_\text{Radio}\approx10^{25}\,$erg\,s$^{-1}$\,Hz$^{-1}$.
The moderate-LBV progenitor on the other hand is about 2 orders of magnitude dimmer than the population average and reaches approximately the same peak-luminosity as the extreme mass-loss case.   

Our results show that the magnetic-field amplification by CR-streaming is sufficient to reproduce the observed radio luminosities assuming electron injection that is congruent with the injection-fraction seen in historical remnants.
The magnetic field in the immediate downstream of the shock peaks after $\approx8\,$days for both progenitors, where the field reaches $\approx8\,$G for the LBV and $\approx1.5\,$G for the RSG progenitor. Afterwards, the field is decaying with a rate of $\propto t^{-1}$. The value of $5\,$G is surpassed after $\approx18\,$days in the LBV-scenario, long before any radio emission could be detected.
However, a contribution of additional magnetic-field amplification in the downstream can not be ruled out.
\cite{1982ApJ...259..302C} proposed that a proportion of the order of a few percent of the shock kinetic energy, converted to magnetic field in the downstream of the shock, would be sufficient to explain the radio emission from very young SNRs. In our case the energy-density of the downstream field constitutes only $\approx0.75\,$\% of the the kinetic energy in the immediate downstream. If this fraction is increased by a factor of 5-10 by additional instabilities in the downstream, the radio luminosity would increase to reach or exceed the observational peak-luminosities.
A similar contribution of post-shock amplification was proposed to be necessary for the youngest known Galactic SNR G1.9+0.3 \citep{2019A&A...627A.166B}.

The evolution of the radio spectral-index\footnote{Assuming a Radio photon-flux $F\propto\nu^\alpha$.} $\alpha$ is shown in the lower panel of Figure \ref{fig:Radio}. The plot indicates that that the canonical DSA spectral index of $\alpha=-0.5$ is reached after the CSM became mostly transparent for the radio emission. 


\section{Conclusions}
\label{sec:conclusions}
We performed numerical simulations of particle acceleration in very young SNRs expanding in dense circumstellar media, solving time-dependent transport equations of CRs and magnetic turbulence in the test-particle limit alongside the standard gas-dynamical equations for CC-SNRs. We derived the CR diffusion coefficient from the spectrum of magnetic turbulence that evolves through driving by the CR-pressure gradient, approximating the non-resonant and resonant streaming instabilities, as well as cascading and wave damping.

We found a good consistency between the thermal X-ray emission predicted from our hydro-models with observations of Type-IIn and Type-IIP SNe with the internal (unabsorbed) luminosities of $\approx10^{43}\,$erg\,s$^{-1}$ and $\approx10^{40}\,$erg\,s$^{-1}$ for Type-IIn and Type-IIP SNe respectively.

The maximum proton energy that we observe in our simulations is well below the PeV-domain. About $600\,$TeV are reached in the presence of a strong stellar field from the progenitor star which could be sufficient to explain the proton CR spectrum up to the knee at $700$\,TeV as reported rcently by the ARGO-YBJ collaboration \citep{2015PhRvD..92i2005B}. 
The value is considerable lower at around $70-200\,$TeV if moderate mass-loss rates for the LBV and RSG progenitors are considered or if the ambient field is lower. The growth of turbulence on the largest scales is hindered by the limited growth-time and the fact that cascading becomes more important as soon as $\delta B\geq B_0$ is fulfilled.

The peak luminosity in the gamma-ray domain reaches $6.3\cdot10^{40}\,$erg\,s$^{-1}$ and $7.2\cdot10^{41}\,$erg\,s$^{-1}$ for the $1-10\,$TeV and $1-300\,$GeV energy-bands respectively, for a Type-IIn progenitor, and $2\cdot10^{38}\,$erg\,s$^{-1}$ and $3.5\cdot10^{39}\,$erg\,s$^{-1}$ for a Type-IIP progenitor.
These limits are fully consistent with a non-detection of gamma-rays from nearby SNe with current-generation instruments.
The intrinsic peak-luminosity is reached $\approx12-30\,$days after the explosion on account of the time needed to accelerate enough particles to the relevant energies. Effects of $\gamma\gamma$ absorption strongly affect the VHE-emission and give rise to a second, local peak in the light-curve of Type-IIP SNe on account of the rapidly fading photosphere.

Current and next-generation gamma-ray instruments might only be able to detect very local SNe with typical detection distances from $20\,$kpc (\emph{HAWC}) to $200\,$kpc (\emph{CTA-South}) for a RSG progenitor and $400\,$kpc (\emph{HAWC}) to $3\,$Mpc (\emph{CTA-South}) for a LBV progenitor.

We investigated the radio emission, taking into account the effect of free-free absorption in the ambient medium. 
We find peak luminosities of $L_\text{Radio}\approx3\cdot 10^{25}\,$erg\,s$^{-1}$\,Hz$^{-1}$ consistent with observations of Type-IIn and Type-IIP explosions albeit being at the lower end of the population average.
As expected from our high mass-loss rates, the rise-times of the radio flux are long with values of $\approx560\,$days and $\approx5600\,$days for a RSG and LBV progenitor respectively.
Using more moderate mass-loss rates gives values of $\approx140\,$days and $\approx700\,$days, closer to the population averages of Type-IIP and Type-IIn explosions. The low\footnote{Low with respect to the population averages of Type-IIP and Type-IIn explosions.} peak luminosities indicate that there might be additional magnetic-field amplification in the downstream of the shocks, additional to the field provided by the streaming of CRs. 

%








\section*{Data availability}
The data underlying this article will be shared on reasonable request to the corresponding author.
\section*{Acknowledgements}
RB and JM acknowledge funding from an Irish Research Council Starting Laureate Award. 
JM acknowledges funding from a Royal Society-Science Foundation Ireland University Research Fellowship.
We acknowledge the SFI/HEA Irish Centre for High-End Computing (ICHEC) for the provision of computational facilities and support (project dsast026c). IS acknowledges support by the National Research Foundation of South Africa (Grant Number 132276).
We are grateful to Morgan Fraser for discussions about the photospheres of IIP and IIn supernovae.

\bibliographystyle{mnras}
\bibliography{References}

\appendix
\section{Details on $\gamma\gamma$-absorption calculation}
\label{ggCalculation}

Unlike \citet{2020MNRAS.494.2760C} where the average optical depth was calculated assuming that the gamma-ray emission is produced in a thin-shell, we calculate the $\gamma\gamma$ opacity, $\tau_{\gamma\gamma}$, as a function of location of the emitted gamma-ray 
This allows us to correctly take into account the complicated distribution of particles both downstream and upstream of the shock when calculating the emission and absorption. We, however, ignore possible light-travel effects that were taken into account in \citet{2020MNRAS.494.2760C} because they play important role only early on, where the internal gamma-ray luminosity is still building up in our simulation.

Exploiting the spherical symmetry assumption we can limit the calculation of the optical depth to an $x$-$y$ plane with $x$-axis directed towards the observer and centered at the center of the explosion,  
\begin{align}
    x &\in [-r_\mathrm{max}, r_\mathrm{max}] \\
    y &\in [0, r_\mathrm{max}] 
\end{align}
with $r_\mathrm{max} = 1.2 R_\mathrm{sh}(t)$, where $R_\mathrm{sh}(t)$ is the shock radius at a certain age.
However, the calculation of the absorption matrices is computationally too expensive to get matrices for every of our output times. Too circumvent this limitation, we produce absorption matrices at 3, 10, 30, 50, 71, 115 and 140 days for the Type-IIP and 5, 12.5, 25 and 100 days for the Type-IIn case and linearly interpolate in time bin-wise between them. To minimize binning effects, the matrices are produced pair-wise with the same $r_\text{max}$. Still, numerical artifacts arise at times where we switch between pairs of matrices that appear as spikes in the gamma-ray light curves.

The estimate of the optical depth is meaningful only within the circle of the radius $r_\mathrm{max}$ as there is no emission produced beyond that.
We also assume that the gamma-ray that travels through the photosphere is completely absorbed.
Hence, we set 
\begin{align}
    & \tau_{\gamma\gamma}\left(\sqrt{x^2+y^2}> r_\mathrm{max}\right) = 0 \\
    &\tau_{\gamma\gamma}\left(x<0; y<R_\mathrm{ps}\right) = \infty
\end{align}

The optical depth is pre-calculated on the spatial grid with $220\times120$ linearly-spaced bins and on the energy grid with 101 logarithmically-spaced bins between 100~keV and 1~PeV. While the energy binning is the same as for radiation calculation the spatial binning is not and further linear interpolation is used on-the-fly. 

The integration over a solid angle in the $\tau_{\gamma\gamma}$ calculation is performed in the spherical coordinate system centered at the location of the emitting region, with zenith in the direction from the centre of the photosphere. With this choice of the coordinate system the integration in zenith angle can be substituted by the integration in $\mu^\prime=\cos{\theta^\prime}$, be limited to $\mu$ from $\mu^\prime_\mathrm{ps} = \cos{\theta_\mathrm{ps}}=\frac{d}{\sqrt{d^2+R_\mathrm{ps}^2}}$ to $1$, where $d$ is the distance between the gamma-ray location and the centre of the photosphere and $\theta_\mathrm{ps}$ is the angular radius of the photosphere as viewed from the gamma-ray location.
Additionally, measuring the azimuth angle from the projected direction towards the observer, we can make use of the symmetry and integrate over $\phi^\prime$ from 0 to $\pi$. Hence, the solid angle integral in Eq.~\ref{eqn:tau_gamma} can be rewritten as
\begin{equation}
    \int_{4\pi}  {\rm d}\Omega = \int_0^{2\pi}d\phi^\prime\int_0^{\theta_\mathrm{ps}}\sin{\theta^\prime}d\theta^\prime = 2 \int_0^{\pi}d\phi^\prime\int_{\mu^\prime_\mathrm{ps}}^1 d\mu^\prime
\end{equation}
The integral along the line-of-sight is calculated in Cartesian coordinates centred on the centre of the photosphere.
The integration is performed up to 100 photosphere radii.  


\begin{figure}
    \centering
    \includegraphics[width=0.49\textwidth]{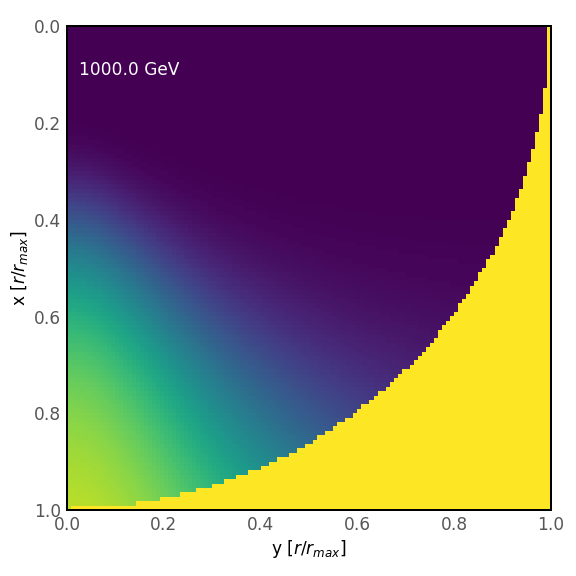}
    \caption{Attenuation factor $\exp(-\tau)$ mapped on the $xy$-plane for two different gamma-ray energies. The colorscale spans from 0 (dark blue) to 1 (yellow).} 
    \label{fig:ggabs}
\end{figure}

Figure \ref{fig:ggabs} shows an example output of the attenuation factor $e^{-\tau_{\gamma\gamma}}$ mapped on the $xy$-plane for two different energies. 
The colorscale spans from 0 (dark blue; complete attenuation of the flux) to 1 (yellow; no attenuation). As described above, simulations are limited to the circle with the radius of $r_\text{max}$ and outside that circle $\tau_{\gamma\gamma}$ is set to zero. For this figure we used a $\delta$-function approximation for the $\gamma\gamma$ cross section \citep{1985ApJ...294L..79Z, 2012rjag.book.....B}
\begin{equation}
    \sigma_{\gamma\gamma}^\delta(\epsilon_\gamma,\epsilon) = \frac{1}{3}\sigma_\mathrm{T}\epsilon_\gamma\delta\left(\epsilon_\gamma - \frac{2}{\epsilon}\right) ,
\end{equation}
which significantly speeds up the simulations does not affect the spatial distribution of the optical depth and hence attenuation factor. The photosphere parameters used here are $R_\text{ps}=5.4\cdot10^{14}\,$cm and $T_\text{ps}=11450\,$K. It can be seen from the figure, that only the emission originating from a cone that is oriented towards the observer is able to reach the observer. The width of this cone depends strongly on the considered gamma-ray energy.

\bsp	
\label{lastpage}
\end{document}